\documentclass[preprint,12pt]{elsarticle}	

\usepackage{amssymb}

\biboptions{sort&compress}

\usepackage{amsmath}

\usepackage[shortlabels]{enumitem}

\usepackage[colorlinks]{hyperref}

\AtBeginDocument{\hypersetup{citecolor=cyan, linkcolor=red,urlcolor=magenta}}

\usepackage{titletoc}

\usepackage{mathrsfs}  

\usepackage{braket}

\def\a{\alpha}
\def\b{\beta}
\def\s{\sigma}

\def\la{\lambda}
\def\La{\Lambda}

\def\ga{\gamma}

\def\vare{\varepsilon}

\def\e{\epsilon}

\def\rm{\mathrm}
\def\cal{\mathcal}

\def\pa{\partial}

\def\be{\begin{equation}}
\def\ee{\end{equation}}
\def\br{\begin{eqnarray}}

\def\er{\end{eqnarray}}
\def\bsub{\begin{subequations}}
\def\esub{\end{subequations}}

\def\Oint{O^{(\rm{int})}}

\journal{Physics Letters B}

\begin{document}

\begin{frontmatter}



\title{Microstate Renormalization in Deformed D1-D5 SCFT}


\author[1]{A.A. Lima}
\ead{andrealves.fis@gmail.com}
\author[1]{G.M. Sotkov}
\ead{gsotkov@gmail.com}
\author[2]{M. Stanishkov}
\ead{marian@inrne.bas.bg}

\address[1]{Department of Physics, Federal University of Esp\'irito Santo, 29075-900, Vit\'oria, Brazil}
\address[2]{Institute for Nuclear Research and Nuclear Energy, Bulgarian Academy of Sciences,
1784 Sofia, Bulgaria}

\begin{abstract}

We derive the corrections to the conformal dimensions of  twisted Ramond ground states in the deformed two-dimensional  $\cal N = (4,4)$ superconformal   $(\mathbb T^4)^N/S_N$ orbifold  theory describing bound states of the D1-D5 brane system in type IIB superstring theory. Our result holds to second order in the deformation parameter, and at the large $N$ planar limit.
The method of calculation involves the analytic evaluation of integrals of four-point functions of two R-charged twisted  Ramond fields and two marginal deformation operators.
We also calculate the deviation from zero, at  first order in the considered marginal perturbation, of the structure constant of the three-point function of two Ramond fields and one deformation operator.

\end{abstract}




\end{frontmatter}



\newpage

\tableofcontents

\section{Introduction}

The bound states of the two-charge D1-D5 brane system describe, at certain limits, an extremal supersymmetric black hole.
Its Bekenstein-Hawking entropy can be calculated by counting BPS states in a particular two-dimen-sional $\cal {N}=(4,4)$  super-conformal  field theory (SCFT$_2$) \cite{Strominger:1996sh,Maldacena:1998bw,Maldacena:1997tm}, while its classical (low-energy) description in IIB supergravity is an asymptotically flat, five-dimensional extremal black hole (or black ring) with a degenerate horizon of radius zero. In the near-horizon (decoupling) limit, this geometry becomes supersymmetric AdS$_3 \times \mathbb S^3 \times \mathbb T^4$, with large Ramond-Ramond charges \cite{Maldacena:1998bw,Seiberg:1999xz}, see also Ref.\cite{David:2002wn} for an extensive review. Mathur's fuzzball proposal \cite{Mathur:2005zp,Skenderis:2008qn} replaces the interior of this extremal black hole (or black ring) with a fuzzy quantum superposition of microstates  described by asymptotically AdS$_3\times \mathbb S^3$ geometries --- including solutions with conical singularities of the form $(\mathrm{AdS}_3\times \mathbb S^3)/Z_N$  in their interiors \cite{Seiberg:1999xz,Balasubramanian:2000rt,Martinec:2001cf} --- and, according to the AdS$_3$/CFT$_2$ correspondence, such geometries can be microscopically expressed in terms of twisted Ramond states of a $\cal N = (4,4)$ SCFT$_2$ with central charge $c = 6 N$. This SCFT$_2$ is best understood at a point in moduli space where, one conjectures, it becomes a free theory with target space $(\mathbb T^4)^N/S_N$; going towards the supergravity limit requires the deformation of this free $S_N$-orbifold theory by a scalar modulus marginal operator \cite{Avery:2010er}.

Extensive research of the orbifold SCFT$_2$ and its deformation has been able to explain essential quantum and thermodynamical properties of black holes 
\cite{Lunin:2000yv,Lunin:2001pw,Balasubramanian:2005qu,Avery:2009tu,Pakman:2009ab,Pakman:2009mi,Burrington:2012yq,Bena:2013dka,Carson:2014ena,Carson:2015ohj,Fitzpatrick:2016ive,Burrington:2017jhh,Galliani:2017jlg,Bombini:2017sge,Tormo:2018fnt,Bena:2019azk,Dei:2019osr,Giusto:2018ovt,Martinec:2019wzw,Hampton:2019csz,Warner:2019jll,Dei:2019iym},
but the complete description of the spectra and the dynamics of  the deformed SCFT$_2$ --- the energies and charges of the fields  and  their multi-point correlation functions --- is  still missing. 
Recent progress 
\cite{Guo:2019pzk,Keller:2019suk,Keller:2019yrr,Guo:2019ady,Belin:2019rba}
in understanding the rules which select protected states from those that get `lifted', i.e. whose  conformal dimensions change, indicates a need for more efficient methods of calculation for the energy lifts of  `$n$-strands'  twisted states.
 
The problem addressed in the present letter  concerns the renormalization of twisted ground-state Ramond fields in the deformed theory.
We present a method for calculating an explicit analytic  expression for the  $\lambda^2$-correction of their conformal dimension,
\be
\Delta^R_n(\la) = \Delta^R_n (0) + \tfrac{1}{2}\pi \lambda^2 |J_R(n)|,	\label{RenormDimens}
\ee 
where $\Delta^R_n(0)$ is the ``bare'' dimension of the Ramond fields in the free orbifold point.
We calculate $J_R$ by integrating a four-point function, and exploring an analytic continuation of `Dotsenko-Fateev' integrals. Our main result is an expression for $J_R(n)$, which is finite, non-vanishing, and at large $n$ seems to stabilize around definite values, as shown in Fig.\ref{UnifiedRamondPlot}.

\section{The $(\mathbb T^4)^N/S_N$ free orbifold and its deformation}


The `free orbifold point' theory is composed of $N$ copies of the $\cal N = (4,4)$ SCFT$_2$, identified under the symmetric group $S_N$, thus forming the orbifolded target space $(\mathbb T^4)^N/S_N$. The central charge of each copy, $c = 6$, results in a total central charge $c_{orb} = 6N$.
We work with Euclidean signature, and on the conformal plane $\mathbb E_2$ or its $\mathbb S^2$ compactification (in opposition to the $\mathbb R \times \mathbb S^1$ cylinder picture).
This orbifold SCFT$_2$ can be formulated in terms of free fields.
There are $4N$ real scalar bosons which can be  organized as $2N$ complex bosons $X^a_I$ and their conjugates $X^{a\dagger}_I$, with $a = 1,2$ and $I = 1, \cdots ,N$. Similarly, the  $4N$ holomorphic and  the $4N$ anti-holomorphic  Majorana fermions can be  combined into
 pairs of complex fermions $\psi^a_I(z)$ and $\tilde \psi^a_I(\bar z)$. We will further use their  bosonized  form   $\psi_I^a(z) = e^{i \phi_I^a(z)}$, $\psi^{a\dagger}_I(z) = e^{- i \phi_I^a(z)}$, in terms of a new set of  chiral bosons $\phi^a_I(z)$.
Similar expressions hold for $\tilde \psi^a_I(\bar z)$.
The twisted boundary conditions, specific for the considered orbifold, are implemented by insertion of `twist operators' $\s_n(z,\bar z)$ \cite{Dixon:1986qv,Lunin:2000yv}. 
The global internal symmetry group of $\mathbb T^4$, $\rm{SO(4)_I} = \rm{SU(2)}_1 \times \rm{SU(2)}_2$, and  the  local R-symmetry group $\rm{SU(2)}_L \times \rm{SU(2)}_R$,  yield a collection of conserved holomorphic (and anti-holmorphic) currents, e.g.~$J^{\pm}(z)$ and $ J^3(z)$ of SU(2)$_L$. These latter R-currents, together with the stress-tensor $T(z)$ and with the supercurrents $G^a(z)$ and $\hat G^a(z)$, which form two doublets of SU(2)$_L$, span the holomorphic sector of the $\cal N =(4, 4)$ superconformal current algebra. 
The anti-holomorphic sector is similarly spanned by $\tilde J^3(\bar z)$, $\tilde J^\pm (\bar z)$, $\tilde G^a(\bar z)$, $\tilde{\hat G}^a(\bar z)$ and $\tilde T(\bar z)$. 

The stress-tensor can be written in terms of the free fields in a point-splitted form 
\be\label{stres}
T(z)= - \tfrac{1}{2}\lim_{w\rightarrow z}
\left[ \pa X^a_I(z) \pa X^{a\dagger}_I(w)+\pa\phi^a_I(z)\pa\phi^a_I(w)+\tfrac{6}{(z-w)^2} \right] ,
\ee
with a sum over $a$ and $I$ left  implicit.  The bosonized  R-current  $J^3(z)$ also has a quite simple form,
$J^3(z) = \frac{1}{2} i \sum_I (\pa \phi_I^1 + \pa \phi_I^2)$.
The (half-integer) eigenvalues $(j^3, \tilde \jmath^3)$ of the zero-modes $J^3_0$ and $\tilde J^3_0$ define the R-charges of  the states of the $\cal N = (4,4)$ algebra, while the eigenvalues of  the stress-tensor(s) zero-modes  $L_0$  and $\tilde L_0$ define the conformal weights $(h,\tilde h)$. The sum  $\Delta= h+\tilde h$   gives the conformal dimension of the states and of the corresponding fields as well.  For example, in each SCFT$_2$ copy, Ramond vacua are defined by spin fields with dimensions $(\frac{c}{24}, \frac{c}{24})$ and, in the holomorphic sector, R-charged vacua correspond to the spin fields $S^\pm(z) = \exp [ \pm \frac{1}{2} i (\phi^1 + \phi^2) ]$, forming a doublet of SU(2)$_L$ with $j^3 = \pm \frac{1}{2}$. More precisely, we will be interested in the R-charged  \emph{twisted} Ramond field $R_{[n]}^\pm(z,\bar z)$, which has twist $n$, R-charges $ (\pm \frac{1}{2}, \pm \frac{1}{2})$ and  conformal weights $(\frac{n}{4},\frac{n}{4})$. Its holomorphic part can be written as
\be\label{tw-ramond}
R_{[n]}^\pm(z)= {1\over\sqrt {n N! (N-n)!}}\sum_{h\in S_N} e^{\pm{i\over 2n}\sum_{I'=(h(1) \cdots h(n))}(\phi_{I'}^1+\phi_{I'}^2)}\sigma_{h^{-1}(1\cdots n)h}.
\ee
Brackets around the twist  $[n]$ will  indicate an $S_N$-invariant combination of  length-$n$ single-cycle twists, obtained by summing over the elements of the conjugacy class of $(1\cdots n) \in S_N$, as in the r.h.s.~of (\ref{tw-ramond}), where the combinatorial factor ensures proper normalization, see \cite{Pakman:2009ab, big_MAG}.
Below, we always assume that $2 \leq n \leq N$.

One moves away from the free orbifold point, and towards the supergravity%
\footnote{Or eventually to certain other limits of  AdS$_3 \times \mathbb S^3 \times \mathbb T^4$  sigma model, for example to  the singular locus in the moduli space of D1-D5 system \cite {Seiberg:1999xz}.}   description, with a deformation,
\be
S_{def}(\lambda)=S_{orb} + \lambda \! \int d^2z \, \Oint_{[2]}(z,\bar z), 	\label{def-cft}
\ee
where $\lambda$ is a dimensionless coupling constant. 
The scalar modulus interaction operator $\Oint_{[2]}(z, \bar z)$ is marginal, with conformal dimension $\Delta_{\rm{int}}=h_{\rm{int}} + \tilde h_{\rm{int}} = 2$, and  protected  from renormalization. It is a singlet of the R-symmetry group, and constructed from NS modes of the supercharges, 
\be\label{def-oper}
\Oint_{[2]} (z, \bar z) = \left( \hat G^1_{-1/2}\tilde G^2_{-1/2}- G^2_{-1/2}\tilde{\hat G}^1_{-1/2}\right) O_{[2]} (z,\bar{z})+ c.c.
\ee
where $O_{[2]}$ is a chiral primary NS field with twist 2, conformal weights $(h, \tilde  h) = (\frac{1}{2},\frac{1}{2})$  and R-charges $(j^3, \tilde \jmath^3) = (\frac{1}{2},\frac{1}{2})$, see e.g. \cite{Lunin:2000yv}. 


\section{Four-point functions and renormalization of Ramond fields}

The effect of the deformation (\ref{def-cft}) on the conformal data of fields is described by conformal perturbation theory \cite{big_MAG}. In this letter we are interested in the changes to the conformal dimension of twisted  Ramond fields $R^\pm_{[n]}(z, \bar z)$ of conformal dimension $\Delta^R_n(0) = \frac{n}{2}$ at the free orbifold point \cite{Lunin:2000yv,big_MAG}.
In the deformed theory (\ref{def-cft}), the dimension becomes a function $\Delta^R_n(\la)$, which can be determined order-by-order in the parameter $\la$ by looking at the corrections to the two-point function $\langle R^-_{[n]}(z_1,\bar z_1) R^+_{[n]}(z_2, \bar z_2) \rangle_\la$. At first order, the change is proportional to the integral of the 3-point function 
$\langle R^-_{[n]}(\infty) \Oint_{[2]}(z )R^+_{[n]} (0)\rangle_0$. 
This function  however vanishes, since  there is no field $R^+_{[n]} $  in the OPE $\Oint_{[2]}(z) R^+_{[n]}(0)$ at all  \cite{big_MAG}.

At second order in $\la$, the correction to the two-point function is given by the integral
\be
\tfrac{\la^2}{2} \!\! \int \!\! d^2z_2 \!\! \int \!\! d^2z_3 
 \big\langle R_{[n]}^-(z_1, \bar z_1) \, \Oint_{[2]}(z_2,\bar{z}_2)  \, \Oint_{[2]} (z_3,\bar{z}_3) \, R_{[n]}^+(z_4, \bar z_4) \big\rangle, 
	\label{second-cor}
\ee
with the $S_N$-invariant  four-point function  evaluated at the free orbifold point.
Conformal invariance implies that
\be
\begin{split}
& \Big\langle R^-_{[n]} (z_1, \bar z_1 ) \Oint_{[2]} (z_2,\bar{z}_2) \Oint_{[2]} (z_3,\bar{z}_3) R^+_{[n]} (z_4, \bar z_4) \Big\rangle 
	=\frac{G(u, \bar u)}{|z_{13}|^4|z_{24}|^{4} |z_{14}|^{n-4}},
\end{split}		 \label{4-point-R-1}
\ee
where $G(u,\bar{u}) = G(u) \bar G(\bar u)$ is an arbitrary function of the anharmonic ratio $u = (z_{12}z_{34})/(z_{13}z_{24})$.	
Global SL(2,$\mathbb C$) transformations can be used to fix $z_1=\infty$, $z_2=1$ and $ z_4=0$; as a consequence, $u=z_3$ and 
\be
 G(u) = \big\langle R^-_{[n]} (\infty) \Oint_{[2]} (1) \Oint_{[2]} (u) R^+_{[n]} (0) \big\rangle .
	\label{gu}
\ee

The standard technology for calculation of multi-point functions in the orbifolded theory is the `covering surface technique' of Lunin and Mathur \cite{Lunin:2000yv,Lunin:2001pw}. Applied to a four-point function, the idea is to map the `base sphere' $\mathbb S^2_{\rm{base}}$, with the four twist operators (inserted at the branching points on $\mathbb S^2_{\rm{base}}$), to a `covering surface' $\Sigma_{\rm{cover}}$, on which the twist operators are trivialized and one is left with a free $\cal N = (4,4)$ SCFT$_2$ without any twisted sectors. 
For large $N$, we can consider $\Sigma_{\rm{cover}}$ to have genus ${\bf g} = 0$, i.e. to be a `covering sphere' $\mathbb S^2_{\rm{cover}}$ \cite{Lunin:2000yv,Pakman:2009zz}.
For the twists in (\ref{gu}),  the unique map $\mathbb S^2_{\rm{cover}} \mapsto \mathbb S^2_{\rm{base}}$ is known to be \cite{Lunin:2000yv,Lunin:2001pw,Pakman:2009zz} 
\be\label{cover}
z(t)=\left({t\over t_1}\right)^n \left( \frac{t-t_0}{t_1-t_0} \right) \left( \frac{t_1-t_\infty }{t-t_\infty} \right),
\ee
where $z \in \mathbb S^2_{\rm{base}}$ and $t \in \mathbb S^2_{\rm{cover}}$. If we label the image of $u$ on the covering by $x$, such that $z(x) = u$, then the correct monodromy requires that the parameters $t_0$, $t_1$, $t_\infty$ must be functions of $x$, which defines a function $u(x)$. 
%
%
With the parametrization
\be
t_0 = x -1,
\quad
t_\infty = x - \frac{x}{x+n} ,
\quad
t_1 = \frac{t_0 t_\infty }{x}
	\label{ArtFrolChoice}
\ee
we obtain the `Arutyunov-Frolov map' \cite{Arutyunov:1997gt},
\be
u(x)={x^{n-1}(x+n)^{n+1}\over (x-1)^{n+1}(x+n-1)^{n-1}} .
	\label{ux}
\ee

%
%
%
\begin{figure*} 
\centering
\includegraphics[scale=0.4]{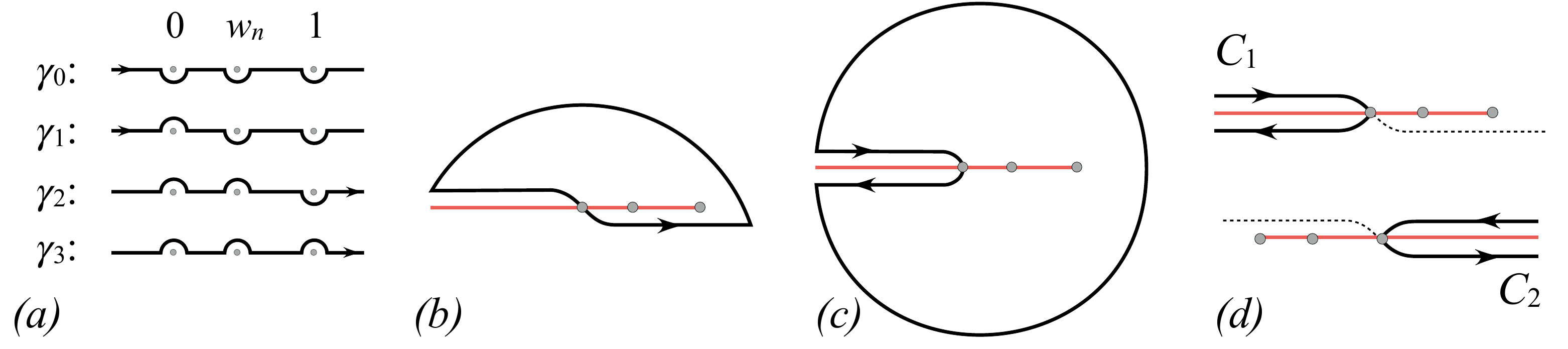}
\caption{
\textit{(a)} Contours for the Dotsenko-Fateev integral;
\textit{(b)} Closing $\gamma_1$;
\textit{(c)} Deformation;
\textit{(d)} Final contours (ignoring circles at infinity). Red lines indicate branching cuts.}
\label{ContoursAndDeformation}
\end{figure*} 
%
%
%

To calculate (\ref{gu}), we use  the `stress-energy tensor method' pioneered in \cite{Dixon:1986qv}, cf. also  \cite{Pakman:2009ab,Pakman:2009mi,big_MAG}. For this we must compute the pole term of the auxiliary function 
\begin{align}
\label{AuxFuncFH}
\begin{split}
	 \frac{\big\langle T(z) R^-_{[n]}(\infty) \Oint_{[2]}(1) \Oint_{[2]} (u , \bar u) R^+_{[n]}(0) \big\rangle}{\big\langle R^-_{[n]}(\infty) \Oint_{[2]}(1) \Oint_{[2]} (u , \bar u) R^+_{[n]}(0) \big\rangle} 
	= 
	 \frac{H(u)}{z - u}  + \cdots 
\end{split}
\end{align}
whereas a Ward identity yields the differential equation
\be\label{equ}
\pa_u \log G(u )= H(u) ,
\ee 
with a similar anti-holomorphic counterpart for $\bar G(\bar u)$. 

The crucial step is to find  $H(u)$. One first lifts the l.h.s.~of (\ref{AuxFuncFH}) to the covering surface, where the twists trivialize. Apart from constant factors that cancel, the twisted Ramond fields $R^\pm_n(z,\bar z)$ are mapped to the SU(2)$_L$ doublet of spin fields:
$R_n^{\pm}(z) \mapsto S^\pm(t)$ inserted \emph{on the covering}. 
Also on the covering, the deformation operator can be expressed as a sum of terms, each a product containing bosonic operators $\pa X^a(t)$ or $\pa X^{a\dagger}(t)$ and ``fermionic exponentials'' $\exp [ \pm\frac{i}{2} (\phi^1(t) + \phi^2(t)]$;   see e.g.~ \S2.3 of \cite{Burrington:2012yq} or \cite{big_MAG}. It is then not difficult to compute the equivalent of (\ref{AuxFuncFH}),
\be
\begin{split}
	& \frac{\big\langle T(t) S^-(\infty) \Oint(t_1) \Oint (x , \bar x) S^+(0) \big\rangle}{\big\langle S^-(\infty) \Oint(t_1) \Oint(x , \bar x) S^+(0)  \big\rangle}
\\
	&= \frac{(t_1 - x)^2}{( t - t_1 )^2 (t - x)^2} 
	 - \frac{1}{4} \left[ \frac{1}{t^2} + \left( \frac{1}{t-t_1} -  \frac{1}{t-x} \right)^2  \right].
\end{split}		\label{FcoverRamond}
\ee
Note that, by construction, on the covering we do not have copy indices $I$, nor the twist label in $\Oint$. The first term in the r.h.s.~of (\ref{FcoverRamond}) comes from contractions of the bosonic part of $T(t)$ with the $\pa X^a$ bosons in $\Oint$; the second term comes from contractions of the fermionic part of $T(t)$ with the fermions in $\Oint$ and $S^\pm$; the spin fields contribute only to the term $1/t^2$ inside the square brackets. We note that after computing the contractions in the numerator of (\ref{FcoverRamond}), one must ``reconstruct'' the four-point function in the denominator, which cancels leaving only the expression in the r.h.s.%
\footnote{The detailed description of all  steps in these calculations will be presented  in our forthcoming paper \cite{big_MAG}.}

%
%

Going back to $\mathbb S^2_{\rm{base}}$ we must also take into account the transformation of $T(t)$, to find that the l.h.s.~of (\ref{AuxFuncFH}) is
\begin{align}
\label{FzusumovaT}
\begin{split}
2 \Bigg[ \frac{c}{12} \big\{ t, z \big \} &+ \left(\frac{dt}{dz}\right)^2  \frac{(t_1 - x)^2}{\left( t(z) - t_1 \right)^2 \left( t(z) - x \right)^2} 
\\
	 &- \frac{1}{4} \left( \frac{dt}{dz}\right)^2   \left( \frac{1}{t^2(z)} + \left( \frac{1}{ t(z) - t_1 } -   \frac{1}{ t(z) - x } \right)^2 \right) \Bigg] ,
\end{split}
\end{align}
where $\{t,z\}$ is the Schwarz derivative coming from the anomalous transformation of the stress-tensor. In (\ref{FzusumovaT}), the function $t(z)$ is one of the two maps obtained by locally inverting (\ref{cover}) near the point $z \approx u$, which must be done by power series expansions, see \cite{Pakman:2009mi,Arutyunov:1997gt}; this multiplicity of the map 
$t : \mathbb S^2_{\rm{base}} \to \mathbb S^2_{\rm{cover}}$ 
is responsible for the factor of 2 above. 
Isolating the term $\sim (z - u)^{-1}$ in (\ref{FzusumovaT}) is not actually feasible, because of the presence of $x$, also implicit in the parameters (\ref{ArtFrolChoice}).
In order to express the r.h.s.~of (\ref{FzusumovaT}) explicitly in terms of $u$,  one should then be able to find the functions $x(u)$, i.e.~the multiple inverses of (\ref{ux}). Instead, we make a change of variables from $u$ to $x$, so that Eq.(\ref{equ}) becomes
\be
\pa_x \log G(u(x)) = (du /dx) H(u) |_{u = u(x)},
\ee
and the r.h.s.~is now a very simple function of $x$  \cite{big_MAG}, whose integral gives the 4-point function with two twisted Ramond fields we are interested in:
\be
G(u(x)) = C_R \; \frac {x^{\frac{5(2-n)}{4}}(x-1)^{\frac{5(2+n)}{4}}(x+n)^{\frac{2-3n}{4}}(x+n-1)^{\frac{2+3n}{4}}}{(x+\frac{n-1}{2})^4}.
 \label{func}
\ee
As mentioned, to express this  function $G(x)$  explicitly in terms of $u$, one needs to invert (\ref{ux}), which can be done only locally by expanding the functions around a specific point. An important example is the limit of coincidence between the deformation operators, i.e. $u \to 1$. Inverting (\ref{ux}) near this point and inserting the result into (\ref{func}), we can check \cite{big_MAG}  that $G(u)$ does not lead to the term associated to $\Oint_{[2]}$ in the OPE of two deformation operators, as it was to be expected. Indeed, this calculation gives the same result as recently found in Ref. \cite{Burrington:2017jhh} by taking the limit of  the four-point function with two BPS chiral (twisted) NS fields. This limit also allow us to  fix  the constant $C_R= 1/ 16 n^2$.

While the method of \cite{Burrington:2017jhh} differs from ours, a calculation analogous to the one above has been done in \cite{Pakman:2009mi} for a four-point function similar to (\ref{gu}) but involving, in place of the twisted  Ramond fields, their chiral NS counterparts with twist $n$, R-charge $\frac{1}{2} (n-1)$ and equal conformal dimension. As it is well-known,
these NS fields are related to the Ramond ones by appropriate spectral flow transformations (see, for example \cite{Burrington:2012yq}).  
The difference between the function $G_{\rm{NS}}(x)$ found for the chiral NS fields (cf.~Eq.(D.6) in \cite{Pakman:2009mi}), and our function (\ref{func}) is  in the values of the exponents of the factors in the numerator. But this is a crucial difference: the exponents in $G_{\rm{NS}}(x)$ are integers for any  $n \in \mathbb N$ (but recall  $2 \leq n \leq N$), hence the function does not have the ``square-root branch cuts'' that  appear in $G(x)$.  This difference in the analytic properties of $G(x)$ and $G_{\rm{NS}}(x)$ is responsible for the fact that the integral of $G_{\rm{NS}}(x)$ corresponding to (\ref{second-cor}) vanishes \cite{Pakman:2009mi}, hence there are no second-order corrections  to the two point function of these BPS chiral NS fields.

We can now proceed with the calculation of the second-order correction to  $\langle R_{[n]}^- R_{[n]}^+ \rangle_{\lambda}$,  by inserting $|G(x(u))|^2$  into  (\ref{second-cor}),
\be\label{correct}
\begin{split}
 \frac{\frac{1}{2}\lambda^2}{|z_{14}|^n}\int \! d^2 z_3 \, & \frac{|z_{14}|^2}{|z_{13}|^2|z_{34}|^2} \int \! d^2 u \;  G(u,\bar u) 
= 
  \frac{\lambda^2\pi}{|z_{14}|^n} \log{\Lambda\over  |z_{14}|}  \int \! d^2u \, G(u,\bar u) .
\end{split}\ee
We have used $z_3$ and $u$ as integration variables, and introduced a cutoff $\La$ to regulate the divergent integral over $z_3$. 
The logarithm at the r.h.s. indicates that there will be renormalization of the conformal dimension of $R _{[n]}^\pm$, given by the remaining integral over $u$.

Hence we turn to calculating the latter integral,
\be\begin{split}
J_R &\equiv \int \! d^2u \; G(u, \bar u) 
\\
	&= \int \! d^2x \; |u'(x)G(x)|^2
\\
	&\equiv [n(n+1)C_R]^2 I , 	\label{JandI}
\end{split}\ee
making use of the function $G(x)$. Note that after the change of variables, all we need to know is the function $G(x)$ that we found on the covering surface.
By a  convenient change of the variables,
$y(x) = -\frac{4(x-1)(x+n)}{(n+1)^2}$,
the integral $I$ defined in (\ref{JandI}), becomes
\be
\begin{split}
I = \int \! d^2y \;  |y|^{2a}|1-y|^{2b}|y-w_n|^{2c}, 	\label{hyper}
\quad
w_n = \tfrac{4n}{(n+1)^2} ,
\end{split}
\ee
a double integral over the complex plane studied in detail by Dotsenko and Fateev \cite{Dotsenko:1984nm,Dotsenko:1984ad,dotsenko1988lectures}.
The exponents in (\ref{hyper}) are
\be
a = \tfrac{1}{2} + \tfrac{1}{4}n , \quad  b = - \tfrac{3}{2} ,  \quad c = \tfrac{1}{2} - \tfrac{1}{4}n  \label{abcwRamond}
\ee
so $I$ is clearly divergent at $y = 1$ and $y = w_n$. The integral is also divergent at $|y| = \infty$. 
As we now show, however, $I$ does have a well-defined, finite value, obtained through an analytic continuation.

%
%
\begin{figure*}
\centering
\includegraphics[scale=0.67]{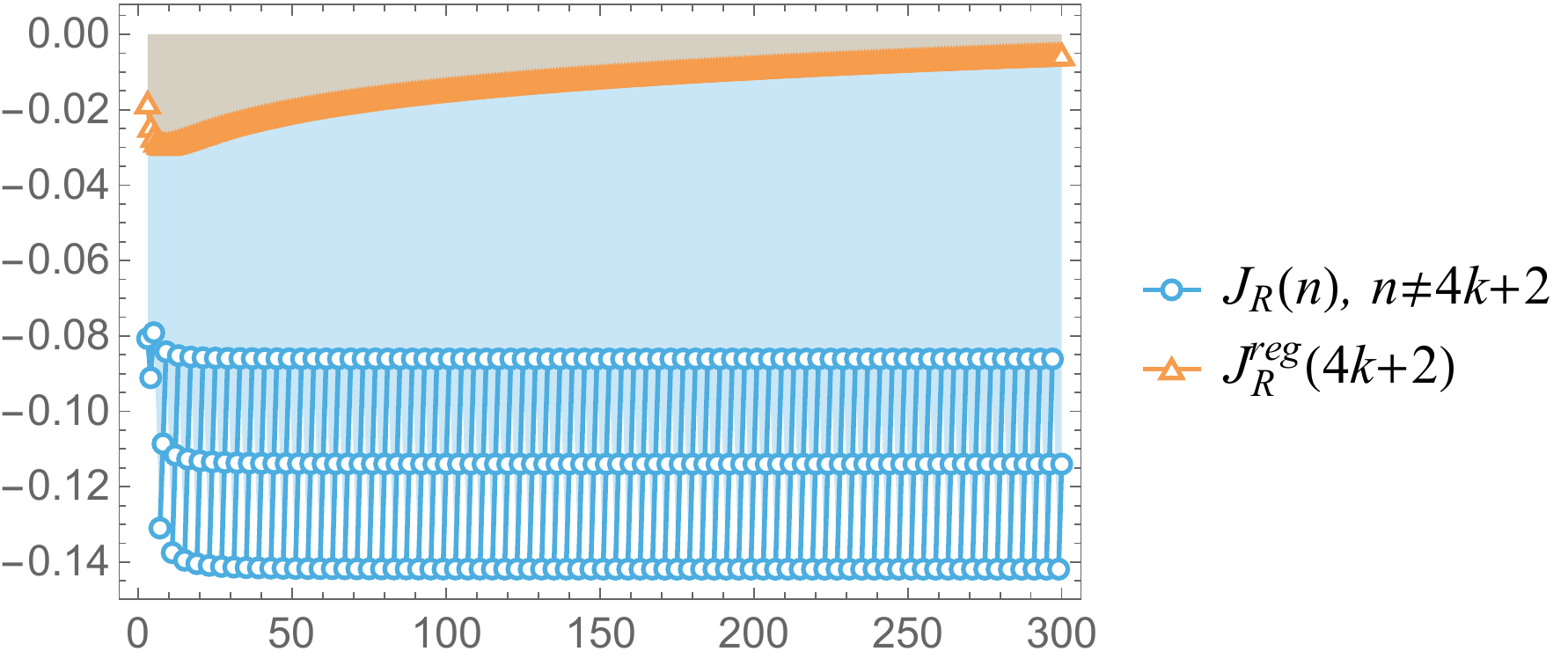}
\caption{The (finite part of the) integral $J_R$ for every $n$. 
The horizontal axis simultaneously denotes two different numbers: $n$ for $J_R(n)|_{n \neq 4k+2}$, and  $k$ for $J_R^{reg}(4k+2)$.}
\label{UnifiedRamondPlot}
\end{figure*} 
%
%
%

In order to solve the Dotsenko-Fateev  integral (\ref{hyper}) we follow \cite{dotsenko1988lectures}, and perform an analytic rotation of the axis $\rm{Im}(y) \mapsto i(1-2 i \vare) \rm{Im}(y)$ with $\vare$ positive and arbitrarily small. The double integral factorizes into 
\be
\begin{split}
I = \frac{i}{2}  \Bigg[  &\int_0^{w_n} \!\! f(v_+)  dv_+  \! \int_{\gamma_1} \!\! f(v_-)  dv_-
				+ \int_{w_n}^1\!\! f(v_+)  dv_+  \! \int_{\gamma_2} \!\! f(v_-)  dv_-
\\
	+  &\int_{-\infty}^0 \!\! f(v_+)  dv_+ \! \int_{\gamma_0} \!\! f(v_-)  dv_-
		+ \int_1^\infty \!\! f(v_+)  dv_+ \! \int_{\gamma_3} \!\! f(v_-)  dv_- \Bigg] 
\end{split}
\ee
where $f(\zeta) =  \zeta^a (\zeta - 1)^b (\zeta - w_n)^c$ and the way the contours $\ga_k$ go around the branch points $0, w_n,1$ is  determined by  $\vare$ as shown in Fig.\ref{ContoursAndDeformation}\textit{(a)}. 
These ``unidimensional'' integrals diverge for the values of $a,b,c$ given in (\ref{abcwRamond}), and here starts our regularization procedure. Assume instead that $a,b,c$ are such that the integrals do exist, and are finite at branching points $0,1,w_n$. This means, in particular, that we can deform the contours to pass \emph{on} these branching points, and close them with a semi-circle as shown in Fig.\ref{ContoursAndDeformation}\textit{(b)} for $\ga_1$. Since the integrand is analytic outside of the branching points, the semi-circle can be deformed as in Fig.\ref{ContoursAndDeformation}\textit{(c)}. The same assumption about $a,b,c$ made above implies \emph{also} that the integral over the circle in Fig.\ref{ContoursAndDeformation}\textit{(c)} vanishes for an infinite radius. Then the integrals originally over the contours $\ga_0$ and $\ga_3$ vanish, while the integrals originally over $\ga_1$ and $\ga_2$ become integrals over $C_1$ and $C_2$ in Fig.\ref{ContoursAndDeformation}\textit{(d)}, coasting the branching cuts (in red) and acquiring, each, a phase of the form $s(\theta) \equiv \sin(\pi \theta)$. The final result is that 
\be
I(a,b,c;w_n) = - s(a)  \tilde I_1 I_2 - s(b) I_1 \tilde I_2 \ ,
	\label{Iafetphases}
\ee
where we introduce the `canonical integrals'
\bsub\begin{align}
I_1(a,b,c;w_n) &\equiv \int_1^\infty   z^a (z - 1)^b (z - w_n)^c \, dz 
\\
\tilde I_1(a,b,c; w_n) &\equiv I_1(b,a;c;1-w_n)  
\\
I_2(a,b,c ; w_n) &\equiv \int_0^{w_n}   z^a (z - 1)^b (z - w_n)^c \, dz 	
\\
\tilde I_2(a,b,c;w) &\equiv I_2(b,a;c;1-w_n) 
\end{align}\label{canonints}\esub

The canonical integrals (\ref{canonints}) are all representations of the hypergeometric function, provided the \emph{same} assumptions made above on $a,b,c$ hold. 
Now the crucial point is that, represented as hypergeometrics, $I_{1,2}(a,b,c;w_n)$ and $\tilde I_{1,2}(a,b,c;w_n)$ are entire functions in the variables $a,b,c$, which are well-defined at the values (\ref{abcwRamond}). Hence the hypergeometric representation is the unique analytic continuation of these functions, and, with Eq.(\ref{Iafetphases}), can be taken as the \emph{definition} of the integral (\ref{hyper}). For $a,b,c$ given in (\ref{abcwRamond}), 
\bsub\begin{align}
I_1(n) &= \frac{\pi (4-n^2)}{ 32} \, w_n^2 \, F ( \tfrac{3}{2},  \tfrac{6+n}{4}  ; 3 ; w_n)
\\
I_2 (n) &= \tfrac{1}{2}  \Gamma ( \tfrac{6 - n}{4} ) \Gamma (\tfrac{6+n}{4} ) w_n^2 F ( \tfrac{3}{2},  \tfrac{6+n}{4}  ; 3 ; w_n)
\\
\tilde I_1(n) &= - 2 \sqrt\pi \, \Gamma( \tfrac{6+n}{4})\,  {\bf F}( \tfrac{n-2}{4} , - \tfrac{1}{2} ;  \tfrac{n+4}{4} ; 1 - w_n )
\\
\tilde I_2(n) &= - \frac{2 \sqrt\pi}{(1-w_n)^\frac{n}{4}} \, \Gamma(  \tfrac{6-n}{4} )  \,  {\bf F}(- \tfrac{n+2}{4} , - \tfrac{1}{2} ;  \tfrac{4-n}{4} ; 1-w_n )	\label{tildI2odd}
\end{align}\label{CanIinteHyper}\esub
where ${\bf F} (\a,\b;\ga;\zeta) \equiv F(\a,\b;\ga;\zeta) / \Gamma(\ga)$, see \cite{NIST:DLMF151}.
We can now evaluate $I$ given by Eq.(\ref{Iafetphases}), then finally evaluate $J_R$ from Eq.(\ref{JandI}),
\be
J_R(n) = - \left( \frac{n+1}{16 n} \right)^2 \Big[ \cos \left(\frac{n \pi}{4}\right) \tilde I_1(n) I_2(n) + I_1(n) \tilde I_2(n) \Big] .
\ee 
The final result, which is finite and non-vanishing, is plotted in Fig.\ref{UnifiedRamondPlot}.
When $n = 4k +2$, a pole of the Gamma function appears in $I_2$ and $\tilde I_2$. One can regularize this Gamma function by taking $k + \e$ with $\e \to 0$, and isolating the singularity in $\Gamma(1-k-\e)$ in a way that is typical of dimensional regularization in QFT, obtaining a finite and an infinite part. The latter has to be renormalized away, see \cite{big_MAG}. The finite result, after some manipulation which relates $\tilde I_1$ to $s(a) I_2$, can be expressed as
\begin{align}
J_R^{reg}(k) &= - \left( \frac{3 + 4k}{32(1 + 2k)} \right)^2 \ I_1(k) \big[ \tilde I_1(k)  +   \tilde I_2^{reg} (k)	\big] 	, \label{FineiJRk}
\\
\begin{split}
\tilde I_2^{reg} (k) &= \frac{(-1)^{k-1} 2 \sqrt\pi (4 k+3)^{2 k+1} \psi(k) }{(4 k+1)^{2 k+1} (k-1)!}  \, 
	{\bf F} \left(-\tfrac{1}{2} , -k-1 ;\tfrac{1-2k}{2} ; \tfrac{(4 k+1)^2}{(4 k+3)^2}\right) ,	\label{FineitilIk}
\end{split}
\end{align}
and is also plotted in Fig.\ref{UnifiedRamondPlot}. (Note that the argument of $J^{reg}_R(k)$ is taken to be $k = \frac{1}{4}(n -2)$, instead of $n$.)
We should make a comment about the limit of large $n$. As seen in Fig.\ref{UnifiedRamondPlot}, the expressions for $J_R(n)$  ``stabilizes'' around finite values. For example, for large $k \in \mathbb N$, we have
$J_R(4(k+1)) \approx -0.1140$, $J_R(2k+1) \approx -0.0861$, $J_R(2k+3) \approx -0.1419$. It is, however, hard to find an analytic expression for these limits, since $n$ enters the hypergeometric functions (\ref{CanIinteHyper})-(\ref{FineitilIk}) in a complicated way.
The final step in the renormalization procedure is to cancel the logarithmic divergence in Eq.(\ref{correct})  by  replacing the bare Ramond fields with their renormalized counterparts  $R^{\pm(ren)}_{[n]} = \Lambda^{ \tfrac{1}{2}\pi \lambda^2 |J_R(n)|} R^{\pm}_{[n]}$. One can easily verify  that  the conformal dimension of the field $R^{\pm(ren)}_{[n]} $   (at $\la^2$-order and in the planar  large $N$ approximation)   is indeed  given by Eq.(\ref{RenormDimens}). 

We have thus found the renormalization of the anomalous dimension (\ref{RenormDimens}) of twisted  Ramond fields $R^\pm_{[n]}$ together with the $\lambda^2$-correction to its two-point function in the deformed orbifold SCFT$_2$ (\ref{def-cft}). 
Our method consisted of a regularization procedure of Dotsenko-Fateev integrals (\ref{hyper}) by analytic continuation, which allowed us to express them in terms of well-defined, finite hypergeometric functions. An obvious check of the validity of this method is to apply it to chiral NS fields --- since these are BPS-protected, the corresponding integral $J_{NS}$ should vanish for all $n$. We can check that this is indeed true; in this case, the integral (\ref{hyper}) is much simpler, and has been discussed in \cite{Pakman:2009mi}.

The same integral $J_R$ which gives the second-order correction of $\Delta^R_n$, also gives the \emph{first} order correction of the specific structure constant $C_n(\la)$ in the three-point function 
$\langle R^-_{[n]}(z_1, \bar z_1) \Oint_{[2]}(z_3, \bar z_3) R^+_{[n]}(z_2, \bar z_2) \rangle_\la$. At zero order, i.e. in the free orbifold, $C_n(0) = 0$, but its correction can be easily calculated to be
$C_n(\la) = \la J_R(n)$, see \cite{big_MAG}, hence
\be
\big\langle R^-_{[n]}(\infty)) \Oint_{[2]}(1) R^+_{[n]}(0) \big\rangle_\la
	=  \la |J_R(n)|  + \cdots,
\ee	
where the ellipsis indicate terms of higher order  in $\la$.

\section{Conclusion}

In this letter we have studied a  simple example of renormalization in the Ramond sector of the deformed orbifold SCFT$_2$ (\ref{def-cft}).
We consider this to be a hint  that  correlation functions involving two generic products of (composite)  twisted Ramond fields (as well as of some of their descendants), and two deformation operators, can be studied with the very same methods used here. The knowledge of the explicit covering surface map seems to be sufficient  for obtaining important information  about the deformed orbifold D1-D5 SCFT$_2$, and consequently  for a more complete microstate description of the related near-extremal 3-charge black holes as well.

\section*{Acknowledgements}

\noindent
The work of M.S. is partially supported by the Bulgarian NSF grant KP-06-H28/5 and that of M.S. and G.S. by the Bulgarian NSF grant KP-06-H38/11.
M.S. is grateful for the kind hospitality of the Federal University of Esp\'irito Santo, Vit\'oria, Brazil, where part of his work was done.
The authors would like to thank an anonymous referee for constructive comments and suggestions.

\section*{References}

\bibliographystyle{elsarticle-num}

\bibliography{D1D5ShortReferencesUpgradeFinal} 

\end{document}